\documentclass[sigconf]{acmart}

\AtBeginDocument{%
  \providecommand\BibTeX{{%
    \normalfont B\kern-0.5em{\scshape i\kern-0.25em b}\kern-0.8em\TeX}}}

\setcopyright{acmlicensed}
\copyrightyear{2018}
\acmYear{2018}
\acmDOI{XXXXXXX.XXXXXXX}

\acmConference[Conference acronym 'XX]{Make sure to enter the correct
  conference title from your rights confirmation emai}{June 03--05,
  2018}{Woodstock, NY}
\acmISBN{978-1-4503-XXXX-X/18/06}

\begin{document}

\title{Exploring the Impact of AI-generated Image Tools on Professional and Non-professional Users in the Art and Design Fields}

\author{Yuying Tang}
\affiliation{%
  \institution{Tsinghua University and Polytechnic University of Milan}
  \city{Beijing and Milan}
  \country{China and Italy}}
\email{tyy21@mails.tsinghua.edu.cn}

\author{Ningning Zhang}
\affiliation{%
  \institution{Tsinghua University}
  \city{Beijing}
  \country{China}}
\email{znn18@tsinghua.org.cn}

\author{Mariana Ciancia}
\affiliation{%
 \institution{Polytechnic University of Milan}
 \city{Milan}
  \country{Italy}}
\email{mariana.ciancia@polimi.it}
 
\author{Zhigang Wang}
\authornote{Zhigang Wang is the corresponding author.}
\affiliation{%
  \institution{Tsinghua University}
  \city{Beijing
}
  \country{China}}
\email{wangzhigang@mail.tsinghua.edu.cn}

\renewcommand{\shortauthors}{Tang et al.}

\begin{abstract}
The rapid proliferation of AI-generated image tools is transforming the art and design fields, challenging traditional notions of creativity and impacting both professional and non-professional users. For the purposes of this paper, we define "professional users" as individuals who self-identified in our survey as “artists,” “designers,” “filmmakers,” or “art and design students,” and "non-professional users" as individuals who self-identified as “others.” This study explores how AI-generated image tools influence these different user groups. Through an online survey (N=380) comprising 173 professional users and 207 non-professional users, we examine differences in the utilization of AI tools, user satisfaction and challenges, applications in creative processes, perceptions and impacts, and acceptance levels. Our findings indicate persistent concerns about image quality, cost, and copyright issues. Additionally, the usage patterns of non-professional users suggest that AI tools have the potential to democratize creative processes, making art and design tasks more accessible to individuals without traditional expertise. This study provides insights into the needs of different user groups and offers recommendations for developing more user-centered AI tools, contributing to the broader discussion on the future of AI in the art and design fields.

\end{abstract}

\ccsdesc[500]{Human-centered computing~Interaction design process and methods}
\ccsdesc[500]{General and reference~Art and Design}
\ccsdesc[500]{Computing methodologies~Artificial intelligence}

\keywords{Creativity Support, Art and Design, Quantitative Methods}

\received{20 February 2007}
\received[revised]{12 March 2009}
\received[accepted]{5 June 2009}

\maketitle

\section{Introduction and Related Work}

Amidst the growing discourse surrounding AI technologies \cite{rae2024effects, bellaiche2023humans, meggert2024art, bellaiche2023humans, grassini2024understanding} and their rapid advancements \cite{creswell2018generative, wang2017generative}, AI Art and AI-assisted Design have emerged, capturing significant attention and reigniting discussions about the democratization of art and design.

AI Art, defined as artwork created using artificial intelligence systems, challenges traditional concepts of creativity and authorship \cite{bellaiche2023humans, cetinic2022understanding, benedikter2021can, chen2020methodological, zylinska2020ai, jiang2023ai}. Notable controversies include AI-generated paintings being auctioned at high prices \cite{goenaga2020critique} and winning international awards \cite{roose2022ai}. Additionally, a renowned photographer rejected an award for an AI-generated photograph in an international competition \cite{Bisson}, and Hollywood screenwriters have resisted AI-generated scripts \cite{chow2020ghost}. Many artists and researchers are further exploring the support that AI-generated image tools provide for artistic creation \cite{10.1145/3610591.3616427, 10.1145/3610537.3622957, 10.1145/3632776.3632827} and creativity \cite{li2024we, li2023ai, zhou2024understanding}. Meanwhile, the applications of AI-generated image tools have expanded to support various design tasks. Designers use AI generation tools like Midjourney, DALL-E 2, and Stable Diffusion to generate images ranging from realistic scenes to product visuals and abstract art, transforming creative workflows \cite{hwang2022too, freese2023ai, quan2023big, jeon2021fashionq}. Some researchers have proposed guidelines for Human-AI Interaction \cite{amershi2019guidelines, li2023assessing, zheng2022ux}, enriching the research framework and methodologies in this field.

Previous research on democratization has primarily focused on how different systems promote the democratization of technology and data \cite{vyas2023democratizing, light2011democratising, 10.1145/2470654.2481360}, with some applications in political contexts \cite{10.1145/3406865.3418562, 10.1145/3311957.3361852}. In the art and design fields, democratization has been discussed in terms of design thinking \cite{urchukov2020design, lal2020democratization, vardouli2012design}, art and design education \cite{burnard2022transdisciplinarity, nakamura2019progressive, orr2019demoncratizing}, regional art transformations \cite{kurlinkus2018nostalgic, hyysalo2019work, rex2016art}, information visualization and interactive platforms \cite{zallio2021democratizing, walmsley2016arts}, and public design \cite{gazulla2024democratizing, chen2023enhancing, kolarevic2018mass, vickery2011art}. Additionally, these discussions often intersect with the concept of Participatory Design \cite{10.1145/3500868.3559449, 10.1145/3406865.3418588, bjorgvinsson2010participatory}. As the availability and capabilities of AI tools continue to evolve, the democratization of AI \cite{sudmann2019democratization, clough2023democratizing, murphy2023democratize, moreau2019paradigm} and its impact on the art and design fields \cite{lal2020democratization, park2024work, cetinic2022understanding, starly2020democratizing, dibia2018designing} have become focal points of interest within the creative community. However, previous studies have not sufficiently explored the potential impact of AI-generated image tools on the democratization of art and design. 


Therefore, our study raises the question of whether existing AI image generation tools can transform the art and design industries, lower the learning or working barriers in these fields, and enable non-professional users to engage in art and design tasks, thus democratizing the creative field. Specifically, our research focuses on the following key questions:

\begin{enumerate}

\item How do the utilization and application of AI-generated image tools differ between non-professional users and professional users?

\item How do the attitudes towards AI-generated image tools differ between non-professional users and professional users?

\end{enumerate}

By investigating these questions, we aim to understand the usage of AI-generated image tools among different user groups and their potential to democratize the art and design fields. The conclusions of this study offer preliminary guidance and suggestions on the impact of AI-generated image tools on the art and design industries and their role in shaping the future of creative practices.

\section{Methodology}

Our research methodology involves quantitative analysis through an online questionnaire (N=380). The primary aim of the questionnaire is to explore the experiences, perspectives, and insights of individuals with and without backgrounds in art and design in their use of AI-generated image tools. We aim to understand the opportunities and challenges these tools present in the art and design fields and to identify any unmet needs among users.

\subsection{Materials}

The online questionnaire, titled ``AI-Generated Image Tools in Art and Design," is divided into six sections:

\begin{enumerate}

\item General information
\item The use of AI-generated static image tools
\item The use of AI-generated static image tools in art and design
\item The use of AI-generated moving image tools
\item AI-generated moving image tools in the art and design process
\item Attitudes and willingness to use AI-generated tools

\end{enumerate}

\subsection{Procedure}

We utilized an online survey platform to record participants' responses. At the beginning of the survey, participants were informed about their rights, the research procedures, and the study's objectives. After obtaining consent, participants provided basic information (e.g., age range, occupation/profession). Throughout the survey, participants answered questions regarding their use of AI-generated image tools and their related user experiences.

\subsection{Participants}

Participants were recruited through mailing lists and social media. Participation was voluntary, and all participants were entered into a random draw to win gift rewards totaling 50 euros. Our survey included a total of 380 participants. The age distribution was 71\% Gen Z (13 to 26 years old), 16\% Gen Y (27 to 42 years old), and 13\% Gen X (more than 42 years old).

Among them, 173 participants (46\%) self-identified their occupations as ``artists,” ``designers,” ``filmmakers,” or ``art and design students.” For brevity, we term this group of participants as ``professional users” in this paper. The remaining 207 participants (54\%) identified their occupation as ``others,” and are referred to as ``non-professional users” in this paper. This classification allows us to explore the user experiences and feedback of different groups when using AI-generated images. However, it has the limitation of relying on self-reported data, which may not always accurately reflect the participants' actual professions.

Out of the total participants, 159 (42\%) reported having used AI static image generation tools. Among them, 102 participants were professional users, while 57 participants were non-professional users. We asked these 159 users follow-up multiple-choice questions to understand the specific user experience when they are using AI-generated image tools. Participants could select multiple options, reflecting the various ways they integrate AI tools into their workflows. Consequently, the reported percentages represent the frequency of selection among users who have utilized AI image-generation tools, inherently allowing the total percentage to exceed 100\% due to the multifaceted use of AI tools by individual users.

Additionally, we inquired about the use of AI-generated moving image tools. Among the professional users, 24 (13\%) reported using AI-generated moving image tools, while 17 (8\%) of the non-professional users reported usage. Due to the relatively small sample size of users who have used moving image tools (41 participants in total), this study primarily focuses on the use of AI-generated static image tools for detailed analysis.

\section{Result}

\subsection{Utilization of AI Tools}

\subsubsection{Professional Users}

Approximately 59\% of participants with art and design backgrounds reported using AI tools to generate static images, with an average usage frequency of 2.5. Among these, 69\% used Midjourney, 30\% used DALL-E 2, and 25\% used Stable Diffusion. These tools were primarily used to create realistic scenes (66\%), followed by abstract art (41\%). The main application areas were graphic design (50\%), product design (27\%), and web design (20\%).

\subsubsection{Non-professional Users}

Among participants without art and design backgrounds, 28\% reported using AI tools to generate static images, with an average usage frequency of 2.7. The most commonly used tools were Midjourney (49\%), DALL-E (39\%), and Stable Diffusion (30\%). These tools were primarily used to create cartoon characters (39\%), cartoon scenes (33\%), realistic characters (32\%), and realistic scenes (32\%). The main application areas were graphic design (35\%), advertising design (32\%), product design (25\%), and web design (25\%).

\begin{figure}
\centering         
\includegraphics[width=0.49\textwidth]{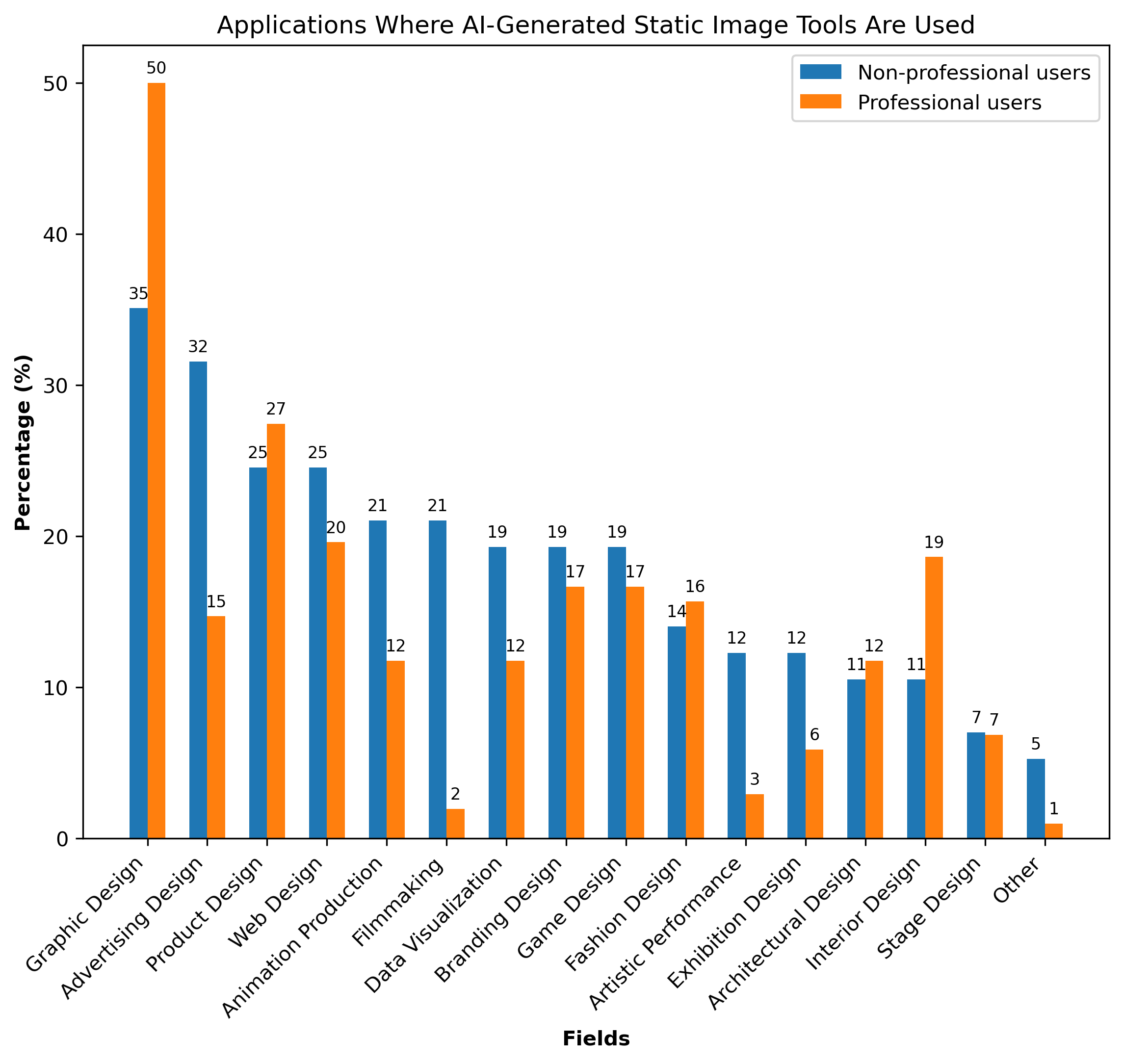} 
 \caption{Applications where Al-generated Static Image Tools are Used}
\label{1}
\end{figure}

\subsubsection{Comparison}

Midjourney emerged as the most popular tool among both groups, indicating its higher user-friendliness. In contrast, the open-source software Stable Diffusion had a higher barrier to entry, reflecting the importance of intuitive interaction and user experience in expanding the user base. Users without an art and design background showed higher usage frequency, suggesting that AI tools are more often integrated into their workflows for art and design tasks. Conversely, professional users relied more on their art expertise and incorporated AI tools into their workflows less frequently. Overall, user frequency of AI tool use was moderately high. In terms of image content, professional users primarily employ AI tools to create images of realistic scenes, followed by abstract art. In contrast, non-professional users more frequently use AI to generate cartoon characters and cartoon scenes, with realistic characters and scenes also being common; the proportions of these categories are relatively balanced. This indicates that non-professional users tend to use AI to produce a more diverse array of image types, whereas professional users show a preference for creating images in realistic styles or more artistically inclined abstract styles. Although both groups primarily used AI-generated images in graphic design, product design, and web design, non-professional users also significantly used them in advertising design, highlighting the value non-professional users place on AI for commercial marketing and promotion (Fig.\ref{1}).

\subsection{User Satisfaction and Challenges}

\subsubsection{Professional Users}

These users generally found the tools easy to use (average ease-of-use rating of 3.1). Common challenges were often resolved through online tutorials and documentation (64\%). The most dissatisfaction was expressed regarding image quality (46\%), followed by price (30\%), and versatility or flexibility of tools (29\%). In contrast, slightly higher satisfaction was noted in the speed of image generation (with only 6\% dissatisfaction) and the quality of user support or tutorials (with only 17\% dissatisfaction).

\subsubsection{Non-professional Users}

Non-professional users also found the tools easy to use (average ease-of-use rating of 3.7). Challenges were often addressed through online tutorials and documentation (61\%). The main areas of dissatisfaction were image quality (42\%), price (35\%), speed of image generation (26\%), and versatility or flexibility of tools (26\%). In contrast, slightly higher satisfaction was noted in the ease of use of the user interface (with only 12\% dissatisfaction).

\begin{figure}
\centering         
\includegraphics[width=0.48\textwidth]{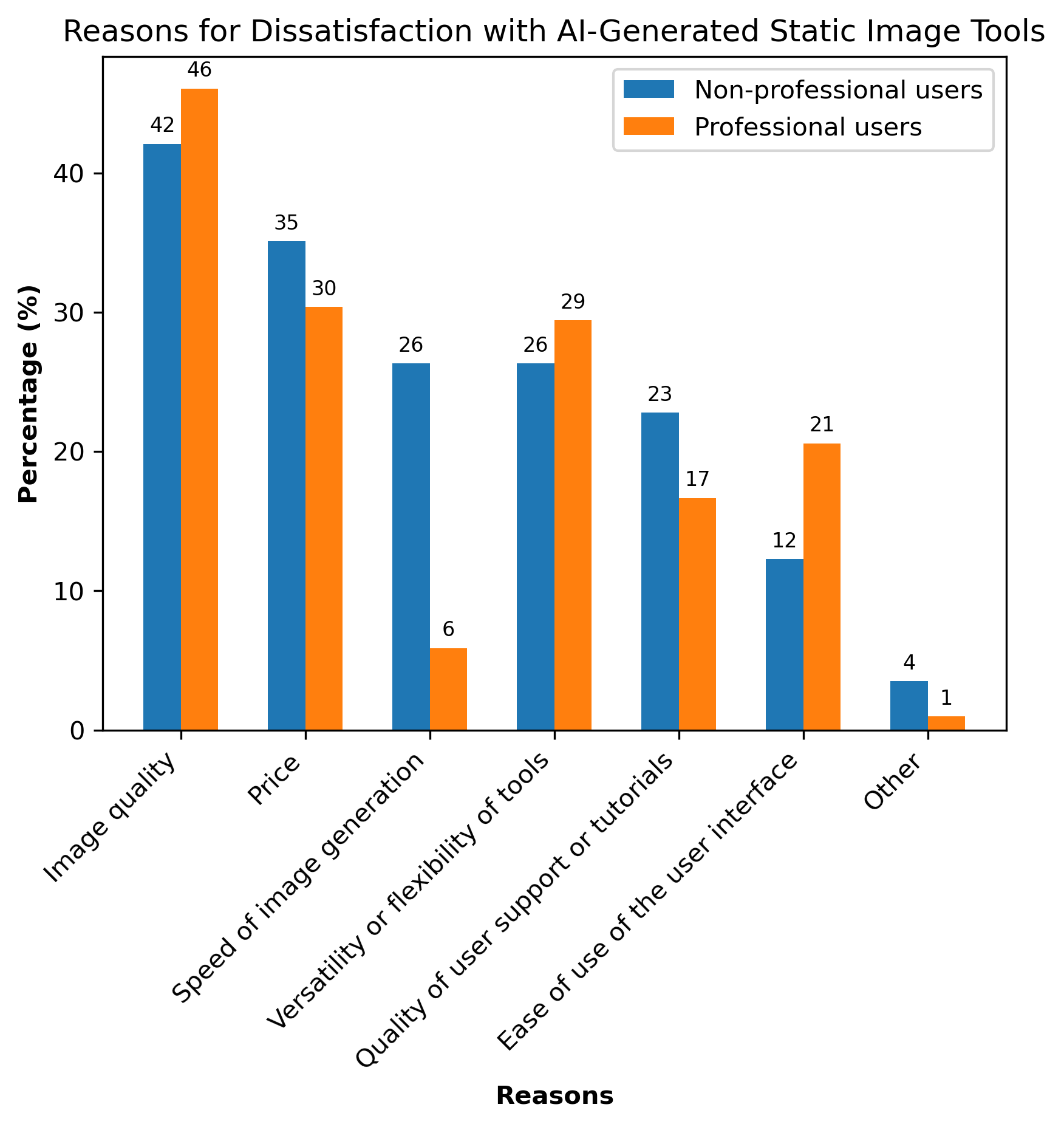} 
 \caption{Reasons for Dissatisfaction with Al-Generated Static Image Tools}
\label{3}
\end{figure}

\subsubsection{Comparison}

Overall, users found AI-generated image tools to be relatively easy to use (on a scale of 1-5). However, dissatisfaction with image quality, price, and versatility or flexibility of tools was common, with professional users being particularly critical of image quality due to their higher standards for detail and precision. Price dissatisfaction was more pronounced among non-professional users, possibly because they perceived the benefits of AI-generated images as lower value.

Meanwhile, non-professional users expressed more dissatisfaction with the speed of image generation, possibly due to their lack of understanding of the time required for traditional creation processes. They directly compare their expectations of AI technology with the actual speed, expecting AI tools to significantly enhance the number of images they can obtain in a fixed time. In contrast, professional users were more satisfied with the speed of AI-generated images, possibly because, for them, AI has already made the creation process much faster than traditional methods. When the frame of reference differs, it is evident that the two groups have distinctly different expectations and levels of satisfaction regarding the speed of AI-generated images.

Professional users expressed more satisfaction with the quality of user support and tutorials, reflecting that they could gain more professional guidance to maximize AI tool use and produce high-quality images. However, non-professional users expressed relatively more dissatisfaction with tutorials, possibly due to the scarcity of tutorials tailored to non-professional users and their lack of knowledge about traditional creation processes, leading to more unmet needs. Nevertheless, non-professional users rated ease of use higher, indicating that they likely use AI tools frequently due to the simple and intuitive interface, which is a crucial attractive element for their experience (Fig.\ref{3}).

\subsection{Application in Creative Processes}

\subsubsection{Professional Users}

AI tools were mainly used in the idea generation stage (65\%) and the actual creation stage (45\%). Specific applications included generating references (72\%) and creating design proposals (38\%). Other commonly used tools alongside AI static image generation tools included ChatGPT (75\%), Photoshop (70\%), Midjourney (51\%), Illustrator (48\%), Figma (42\%), and Blender (22\%).

\subsubsection{Non-professional Users}

Non-professional users also used AI tools primarily in the idea generation stage (61\%) and the actual creation stage (54\%). Specific applications included generating references (58\%), creating design proposals (46\%), and creating 2D moving graphics (40\%). Commonly used tools alongside AI-generated static images included ChatGPT (79\%), Photoshop (39\%), Midjourney (30\%), DALL-E (25\%), Illustrator (21\%), and Stable Diffusion (19\%).

\subsubsection{Comparison}

Overall, users primarily used AI-generated static image tools in the idea generation stage and for generating references and design proposals. However, non-professional users more frequently used AI tools in the actual creation stage, incorporating AI-generated images into their final outputs and extensively using them in 2D moving graphics creation. In contrast, professional users mainly used AI tools for generating references. Non-professional users showed a preference for combining different AI tools, such as DALL-E and Stable Diffusion, while professional users relied more on professional design software like Illustrator, Figma, and Blender. This suggests that non-professional users depend more on various AI tools to achieve results, while professional users leverage their professional skills to refine and enhance AI-generated images.

\subsection{Perceptions and Impacts}

\subsubsection{Professional Users}

Positive impacts included increased inspiration for creative work (54\%), saving time and costs (48\%), and making creation and design more efficient (48\%). Concerns included copyright issues (50\%) and the similarity of generated content (46\%).

\subsubsection{Non-professional Users}

Positive impacts included increased inspiration for creative work (54\%), realizing certain design ideas (54\%), saving time and costs (51\%), and making art and design easier (49\%). Concerns included the similarity of generated content (51\%), reduced human dominance and control in creation (46\%), and lack of innovation and creativity in the generated content (44\%).

\begin{figure}
\centering         
\includegraphics[width=0.49\textwidth]{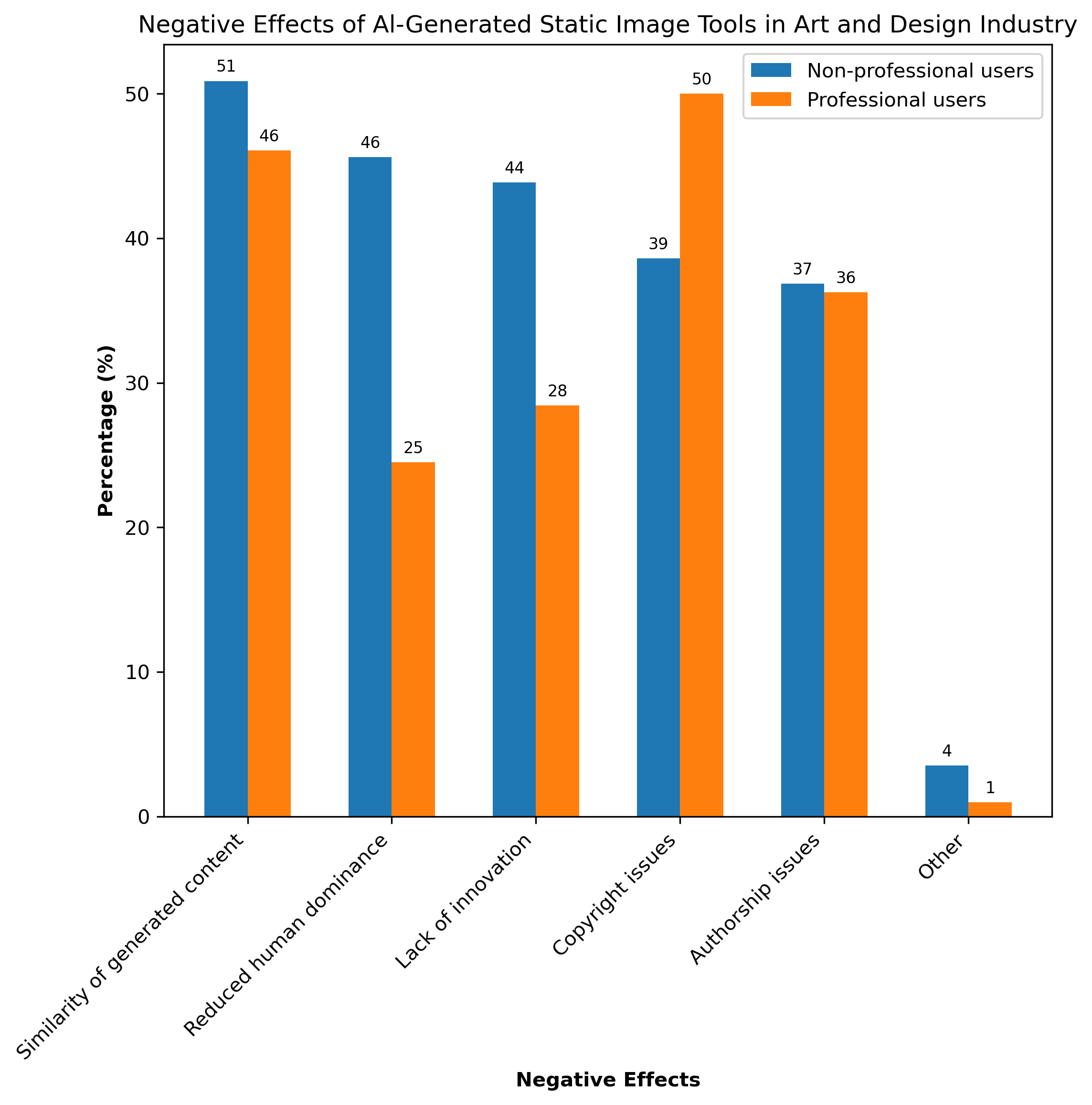} 
 \caption{Negative Effects of Al-generated Static Image Tools in Art and Design Industry}
\label{4}
\end{figure}

\subsubsection{Comparison}

Both groups recognized that AI-generated image tools increased inspiration and saved time and costs, and both were concerned about the similarity of generated content. 

However, during the use of AI tools, non-professional users placed more importance on realizing certain design ideas and making art and design easier. This likely stems from their lack of professional skills in the art and design field, leading them to value the ease of realizing design ideas and creating art and design. Non-professional users were also more worried about reduced human control in creation and the lack of innovation and creativity, whereas professional users felt less concerned, likely due to their ability and expertise to modify AI-generated images more effectively and judge the creativity of AI images. In contrast, professional users were more concerned with copyright issues, reflecting the creative industry professionals' greater focus on the originality and intellectual property of AI-generated images (Fig.\ref{4}).

\subsection{Acceptance Levels}

\subsubsection{Professional Users}

67\% of participants viewed AI tools as just tools, while 33\% saw them as real collaborators (like colleagues, assistants, team members, etc.). Regarding the role of AI, 65\% believed AI aids creativity, 19\% were uncertain, 9\% thought AI might replace human creativity, and 7\% held other views. In terms of acceptance, 45\% had used AI tools and would continue to use them, 36\% had not used but were willing to start, 8\% refused to use them themselves but did not oppose others using them, 6\% had used them but would not continue, and 4\% refused to use them and were strongly against others using them.

\subsubsection{Non-professional Users}

57\% viewed AI tools as just tools, while 43\% saw them as real collaborators. Regarding the role of AI, 66\% believed AI aids creativity, 23\% were uncertain, 9\% thought AI might replace human creativity, and 3\% held other views. In terms of acceptance, 57\% had not used AI tools but were willing to start, 26\% had used them and would continue, 11\% refused to use them themselves but did not oppose others using them, and 3\% each had used but would not continue and refused to use them and were strongly against others using them.

\subsubsection{Comparison}

Both groups generally viewed AI as a means to enhance creativity and mostly saw AI tools as just tools, although non-professional users were more likely to see them as real collaborators. This may be because non-professional users completed tasks with AI that typically required real designers or collaborators. Both groups showed positive and open attitudes toward AI tool acceptance, with most willing to use or try AI tools, and only a small portion strongly opposed to their use.



 


\section{DISCUSSION}

This study investigates the utilization, satisfaction, application, perception, and acceptance of AI-generated image tools among professional and non-professional users. These findings attempt to provide preliminary insights into the role and future development of AI-generated image tools in the creative fields, which can be summarized as follows:

\subsection{Current Usage and Democratizing Potential}

\subsubsection{Enhanced Accessibility}
The higher usage frequency of AI tools among non-professional users suggests that these technologies are lowering the basic barriers to entry for creative activities. Non-professional users can leverage AI tools to perform tasks that traditionally require specialized skills, such as graphic and advertising design. The significant employment of AI tools by non-professional users in commercial areas like advertising design demonstrates the practical applications of these technologies beyond traditional art and design contexts. The application of AI-generated image tools in these fields broadens the scope of who can participate in creative processes and how these tools are utilized, further supporting the democratization of various creative application fields and enabling more individuals to participate in and contribute to creative endeavors.

\subsubsection{Intuitive Interfaces}
The popularity of Midjourney among both groups, particularly non-professional users, highlights the importance of intuitive and user-friendly interfaces in expanding the user base. The higher ease-of-use ratings from non-professional users indicate that simplified interfaces are crucial for making these tools accessible to those without formal training in art and design. This accessibility is a key factor in the democratization process, as it empowers more people to engage in creative endeavors without needing extensive expertise.

\subsubsection{Human-AI Collaboration Enhances the Democratization of AI-generated Content}

The perception of AI tools as collaborators, especially among non-professional users, underscores their transformative impact on traditional creative roles. AI tools enable non-professional users to undertake tasks in art and design that previously required expert knowledge, thereby reshaping the landscape of these fields to be more inclusive and collaborative. This collaborative potential is significant as it allows for a more diverse range of perspectives and ideas to enter the creative process. AI tools can act as equalizers, providing non-professional users with the means to execute complex projects and bring their visions to life. This not only democratizes creativity but also fosters innovation by allowing more voices to contribute to the artistic conversation.

Furthermore, the integration of AI tools can facilitate new forms of collaboration between professional and non-professional creators. Professionals can use these tools to expedite routine tasks, freeing up more time for complex and innovative projects, while non-professionals can leverage AI to participate in high-level creative endeavors. This symbiotic relationship can drive the evolution of art and design, leading to richer and more diverse outputs. By recognizing and harnessing the collaborative potential of AI tools, the art and design fields can become more inclusive and dynamic, embracing a broader spectrum of contributors and fostering a more vibrant creative ecosystem.

\subsection{User-Centred Perspectives on AI Tools}

\subsubsection{User Experience Challenges}
Image quality emerged as a significant area of dissatisfaction, particularly among professional users who require higher standards of detail and precision. Additionally, the cost of these tools was a concern, especially for non-professional users who felt that the benefits did not justify the price. While open-source software like Stable Diffusion can be used for free when deployed locally, its performance is often constrained by the user's hardware capabilities, and the quality of generated images is heavily influenced by the user’s technical knowledge and ability to navigate complex interfaces. Most non-professional users desire a simple process for generating AI images but are concerned about losing too much control over the generation process, which could lead to issues such as excessive similarity in the output. Therefore, there is a need to strike a balance between simplicity and user control to ensure a satisfactory experience.

Future AI-generated image tools have the potential to further democratize the art and design fields, but they must carefully balance image quality and cost, simplicity of interaction, and the degree of user control. If the process is too simple, users might find the results unpredictable and less useful. Conversely, if there are too many controllable parameters, the learning curve might become too steep, discouraging users from adopting the tools.

\subsubsection{Guidelines for User-Centred AI Tools}

To further democratize the art and design fields and enhance user experiences, several improvements are recommended for future AI-generated image tools:

\begin{enumerate}

\item Enhance Image Quality: Prioritize advancements in the detail and precision of AI-generated images to meet the exacting standards of professional users. This could involve integrating more sophisticated algorithms and higher-quality training data to produce images that are more nuanced and realistic. Focusing on these improvements will make AI tools more appealing to users who require high-quality images in their work.

\item Personalize AI Tool Functionality: Tailor AI tools to different user needs by balancing human control, ease of idea realization, generation speed, image quality, and cost. For non-professional users, offer less control with a focus on simplifying prompt writing and quickly generating multiple options for rapid iteration and optimization. For professional users, provide more control parameters to help fine-tune specific ideas or optimize designated areas, ensuring seamless integration with other software in their existing workflows. Develop intuitive interfaces that simplify the generation process without sacrificing essential controls, which could be achieved through guided workflows and customizable templates that offer both simplicity and flexibility. Implement adaptive interfaces that cater to both professional users and non-professional users, such as a basic mode with streamlined options for non-professional users and an advanced mode with more granular controls for professional users.

\item Implement Accessible Pricing Models: Increase adoption rates among non-professional users by offering flexible and affordable pricing options. Consider tiered subscription plans, pay-per-use options, and educational discounts to make these tools more accessible. Lowering the financial barrier will allow users from diverse backgrounds to engage with and benefit from AI-generated image tools, fostering a more inclusive creative community.

\item Improve Educational Resources: Offer tailored educational programs for different types of users, such as advanced courses for professional users and foundational courses for non-professional users to supplement the creative process. Users need strong learning skills and initiative to effectively combine AI tools with traditional creative processes. Expand the availability of high-quality tutorials, user support, and comprehensive training programs to help all users maximize the potential of AI tools. Educational institutions and universities could incorporate courses on using AI in creative processes into their art and design curricula, covering basic usage to advanced techniques. Online platforms could offer specialized workshops and certification programs, ensuring continuous learning opportunities and expert guidance.

\end{enumerate}

By focusing on these areas for improvement, developers can enhance the user experience and broaden the adoption of AI-generated image tools. This will further support the democratization of art and design, allowing a wider range of individuals to engage in and contribute to creative fields.

\subsection{Social and Ethical Considerations}

Both art and non-professional users recognized the benefits of AI tools in increasing inspiration and efficiency. This supports the argument that AI is democratizing creativity by making it more accessible and streamlined. However, concerns raised by users about content similarity, and copyright highlight the need for robust ethical and legal frameworks. Despite platforms like Midjourney offering commercial usage rights within certain limits for paid subscribers, professional users remain cautious about these issues, reflecting broader concerns in the creative industry about the ownership and originality of AI-generated works. Addressing these issues is crucial to ensuring that the democratization process respects intellectual property rights and maintains the integrity of creative work.

Moreover, the reliance on AI-generated content raises ethical questions about the evolving role of human creativity and the balance between human and AI contributions in the creative process. Concerns about the similarity of generated content and the potential reduction of human dominance in creation point to deeper issues about the nature of creativity and innovation in an AI-augmented world. As AI tools become more prevalent, there is a growing need to ensure they enhance rather than diminish the uniqueness and individuality of creative works.

To mitigate these concerns, developers, and policymakers should work together to establish clear guidelines on the use and attribution of AI-generated content. This could include standardized practices for citing AI tools, mechanisms for verifying originality, and legal protections for creators who use AI in their work. By fostering a secure and ethical environment, the creative community can more confidently embrace AI technologies.

\section{Limitation and Future Works}

In the rapidly evolving field of artificial intelligence, the types and usage of AI tools are constantly changing. Our study captures user feedback up to December 31, 2022. Future research should continue to monitor the dynamic changes in user needs to develop more user-centred AI tools. Moreover, with the increasing proliferation of AI tools, it is vital to explore how creative workflows, skill requirements, industry standards, and career development in the art and design field are evolving. Investigating the evolving relationship between human creativity and AI will yield deeper insights into the potential and limitations of AI technologies. It is crucial to further determine whether AI-assisted creativity can truly achieve broader democratization, as this is an essential way to understand the transformative potential of AI in the art and design fields in the future.

Overall, the insights from this preliminary study deepen our understanding of the applications of AI-generated image tools among different user groups in the art and design fields. By continuing to refine AI tools and address the concerns and needs of diverse user groups, we can further explore the potential of AI to foster innovation, inclusivity, and creativity within the art and design industries.

\bibliographystyle{ACM-Reference-Format}
\bibliography{main}

\end{document}